\documentclass[notitlepage,prereprint,12pt]{revtex4-1}

\usepackage{amssymb}
\usepackage{natbib}
\usepackage{graphicx}
\usepackage{amsmath}
\usepackage[bookmarks = false]{hyperref}
\usepackage{bm}%
\usepackage[all]{hypcap}
\usepackage{graphicx}
\usepackage{colortbl}
\usepackage{booktabs}
\usepackage{enumerate}
\usepackage{enumitem}
\usepackage{mathrsfs}
\usepackage{multirow}
\usepackage{upgreek}    
\usepackage{rotating}
\usepackage{setspace}

\newcommand\oprod[2]{\ensuremath{|#1\rangle\langle#2|}}

\usepackage{array}
\newcommand{\PreserveBackslash}[1]{\let\temp=\\#1\let\\=\temp}
\newcolumntype{C}[1]{\rangle{\PreserveBackslash\centering}p{#1}}
\newcolumntype{R}[1]{\rangle{\PreserveBackslash\raggedleft}p{#1}}
\newcolumntype{L}[1]{\rangle{\PreserveBackslash\raggedright}p{#1}}

\begin{document}
\title{Experimental 4-intensity decoy-state quantum key distribution with asymmetric basis detector efficiency}

\author{Hui Liu$^{1,\,2}$}
\author{Zong-Wen Yu$^{3,\,4}$}
\author{Mi Zou$^{1,\,2}$}
\author{Yan-Lin Tang$^{7}$}
\author{Yong Zhao$^{7}$}
\author{Jun Zhang$^{1,\,2}$}
\author{Xiang-Bin Wang$^{3,\,5,\,6}$}
\email{Email Address: xbwang@mail.tsinghua.edu.cn}
\author{Teng-Yun Chen$^{1,\,2}$}
\email{Email Address: tychen@ustc.edu.cn}
\author{Jian-Wei Pan$^{1,\,2}$}
\email{Email Address: pan@ustc.edu.cn}

\affiliation{$^1$Hefei National Laboratory for Physical Sciences at Microscale and Department
of Modern Physics, University of Science and Technology of China, Hefei, Anhui 230026, China}
\affiliation{$^2$CAS Center for Excellence in Quantum Information and Quantum Physics, University of Science and Technology of China, Hefei, Anhui 230026, China}
\affiliation{$^3$State Key Laboratory of Low Dimensional Quantum Physics, Department of Physics, Tsinghua University, Beijing 100084, People's Republic of China}
\affiliation{$^4$Data Communication Science and Technology Research Institute, Beijing 100191, People's Republic of China}
\affiliation{$^5$Jinan Institute of Quantum Technology, SAICT, Jinan 250101, People's Republic of China}
\affiliation{$^6$Department of Physics, Southern University of Science and Technology, Shenzhen, 518055, People's Republic of China}
\affiliation{$^7$QuantumCTek Corporation Limited, Hefei, Anhui 230088, China}

\begin{abstract}
The decoy-state method has been developed rapidly in quantum key distribution (QKD) since it is immune to photon-number splitting attacks. However, two basis detector efficiency asymmetry, which exists in realistic scenarios, has been ignored in the prior results. By using the recent 4-intensity decoy-state optimization protocol, we report the first implementation of high-rate QKD with asymmetric basis detector efficiency, demonstrating 1.9 to 33.2 times higher key rate than previous protocols in the situation of large basis detector efficiency asymmetry. The results ruled out an implicitly assumption in QKD that the efficiency of Z basis and X basis are restricted to be same. This work pave the way towards a more practical QKD setting.
\end{abstract}
\maketitle

\section*{Introduction}
	
Quantum key distribution (QKD) has continuously been focused since the first protocol proposed by Bennett and Brassard in 1984 \cite{bennett1984proceedings}. However, the unconditionally security of the ideal BB84 has been frustrated by a lot of realistic imperfections, one prominent of which is the lack of the practical single photon source. It is more feasible for Alice to utilize the attenuated laser, i.e., the weak coherent pulses (WCP) as signal states, which results in a loophole for the photon-number splitting (PNS) attack\cite{brassard2000limitations,lutkenhaus2000security}. Fortunately, based on the original idea by Hwang \cite{hwang2003quantum}, the decoy-state method\cite{wang2005beating,lo2005decoy} appeared in time. It has dramatically improve the performance of QKD with the attenuated laser by providing better bounds on the gain and the error rate of single photon states. In the past decade, noteworthy theoretical improvements have been proposed to continuously improve the performance of decoy-state QKD\cite{wang2007quantum,hayashi2007upper,jiang2016universally,chau2018decoy}. Experiments either over optical fiber or free-space have advanced significantly in the meantime\cite{rosenberg2007long,schmitt2007experimental,yuan2007unconditionally,chen2009field,liu2010decoy,boaron2018secure,liao2017satellite,peng2007experimental,yin2016measurement,liao2018satellite}. Specially, QKD has been demonstrated at a transmission distance up to 7600~km in free-space\cite{liao2018satellite} and more than 400~km in optical fiber\cite{boaron2018secure,yin2016measurement}.

Nevertheless, the practical applications of QKD combined with the one-time pad scheme are still pinned by low secure key rate. In addition, an implicit assumption for detector model in the existed results is that the efficiencies of Z basis and X basis are almost the same. It seems like a simple assumption but does not always meet realistic scenarios. 
For instance, it could be resulted from the efficiency asymmetry of single photon detectors in the passive basis choice protocol, or the imperfection during measurment bases switching in the active basis choice protocol.
A common approach is reducing higher efficiency to balance detector efficiency asymmetry at the price of introducing additional losses.

Here, by making simple modifications to a commercial QKD system, we implement a novel 4-intensity decoy-state QKD protocol using biased bases\cite{yu2016reexamination}, which can provide higher key rate than previous traditional 3-intensity protocols with unbiased bases, especially in a large degree of basis detector efficiency asymmetry. Setting the detector efficiency asymmetry of two bases $\eta_Z/\eta_X = 2$ and $10$ respectively, we change channel distance over different lengths of standard telecom fiber up to 150~km and demonstrate as much as 1.9 to 33.2 times higher key rate than previous protocols. These results have moved QKD towards a more practical setting.

\section*{Theory}
In the novel 4-intensity QKD protocol\cite{yu2016reexamination}, Alice prepares two different coherent sources in the $Z$ basis with intensities $\mu_{Z_1}$ and $\mu_{Z_2}$; and two different coherent sources in the $X$ basis with intensities $\mu_{X_1}$ and $\mu_{X_2}$, with probabilities $p_{\alpha_j}(\alpha=Z,X;j=1,2)$ respectively. Without losing the generality, we assume $\mu_{\alpha_1}<\mu_{\alpha_2}(\alpha=Z,X)$. The coherent state whose phase is selected uniformly at random can be regard as a mixture of photon number states, i.e., $\rho_{\alpha_j}=\sum_{k} a_{k,\alpha_j}\oprod{k}{k}$ with $a_{k,\alpha_j}=e^{-\mu_{\alpha_j}}\mu_{\alpha_j}^k/k!$ for $\alpha=Z,X$ and $j=1,2$. In the protocol, Bob measures the received pulses in the $Z$ and $X$ bases with probabilities $q^Z$ and $q^X$ respectively. After the preparation and measurement of $N_t$ pulses, Alice and Bob obtain the observable $N_{\alpha_j}^{\omega}$ and $M_{\alpha_j}^{\omega}$ which are the number of successful counts and error counts when Alice sends the pulses from source $\alpha_j$ and Bob measures them in the $\omega$ basis. Here $\alpha$ and $\omega$ can take both $Z$ and $X$. We also denote $S_{\alpha_j}^{\omega}$ and $T_{\alpha_j}^{\omega}$ as the yield and error yield, respectively, with $S_{\alpha_j}^{\omega}= N_{\alpha_j}^{\omega}/(p_{\alpha_j} q^{\omega} N_t)$ and $T_{\alpha_j}^{\omega}= M_{\alpha_j}^{\omega}/(p_{\alpha_j} q^{\omega} N_t)$.

In Ref.~\cite{yu2016reexamination}, a delicate point has been put forward that even in the asymptotic case, i.e., $s_0^{Z}\neq s_0^{X}$ and $s_{1,\alpha_1}^{Z} \neq s_{1,\alpha_2}^{X}$.  Here $s_{k,\alpha_j}^{\omega}$ is the yield of $k$-photon pulses prepared from source $\alpha_j$ and measured in the $\omega$ basis. The reason $s_{1,Z}^{Z}\neq s_{1,X}^{X}$ is simply due to the asymmetry of detection efficiencies and dark counts in different bases. Such asymmetry can come from either imperfect control of two of the devices inside Labs, or Eve's attack. In order to take a better treatment, the decoy-state method jointly in different bases has been studied\cite{yu2016reexamination}. For this goal, the observed number of counts of pulses prepared in one basis but measured in another basis shall be used. In particular, it is assumed that $s_{1,Z}^{Z}=s_{1,X}^{Z}$ and $s_{1,Z}^{X}=s_{1,X}^{X}$ are valid. Given these equations, one does not have to study the decoy-state method completely separately in each basis.

In all real experiment, the total number of pulses sent by Alice is finite. In order to extract the secret final key, we have to consider the effect of statistical fluctuations caused by the finite size. In this case, yields of the same state out of different sources are not always rigorously equal to each other, i.e., $s_{k,\alpha_1}^{\omega}\neq s_{k,\alpha_2}^{\omega}$. Accordingly~\cite{yu2016reexamination}, with the observed values $S_{\alpha_j}^{\omega}$, one can lower bound the mean value $\langle s_{1,\alpha}^{\omega}\rangle$ for a given value of $\langle s_0^{\omega}\rangle$ with the following equations
\begin{equation}\label{eq:Ls1mean}
  \langle s_{1,\alpha}^{\omega}\rangle \geq \langle s_{1}^{\omega,L}\rangle= \max_{\alpha=Z,X} [ \langle s_{1,\alpha}^{\omega,L}\rangle(\langle s_{0}^{\omega}\rangle)],
\end{equation}
and
\begin{equation}\label{eq:s1mean}
  \langle s_{1,\alpha}^{\omega,L}\rangle(\langle s_{0}^{\omega}\rangle) =\frac{1}{A_{\alpha_1 \alpha_2}^{1,2}}\left[ a_{2,\alpha_2} \underline{S}_{\alpha_1}^{\omega} -a_{2,\alpha_1} \overline{S}_{\alpha_2}^{\omega} -A_{\alpha_1\alpha_2}^{0,2} \langle s_0^{\omega}\rangle \right],
\end{equation}
where $A_{\alpha_1 \alpha_2}^{0,2}=a_{0,\alpha_1}a_{2,\alpha_2}-a_{0,\alpha_2}a_{2,\alpha_1}$, $A_{\alpha_1 \alpha_2}^{1,2}=a_{1,\alpha_1}a_{2,\alpha_2}-a_{1,\alpha_2}a_{2,\alpha_1}$, and $\underline{S}_{\alpha_j}^{\omega} =S_{\alpha_j}^{\omega}/(1+\delta_{\alpha_j}^{\omega})$, $\overline{S}_{\alpha_j}^{\omega} =S_{\alpha_j}^{\omega}/(1-\delta_{\alpha_j}^{\omega})$. By using the multiplicative form of the Chernoff bound, with a fixed failure probability $\epsilon$, we can give an interval of $\langle S_{\alpha_j}^{\omega}\rangle$ with the observable $S_{\alpha_j}^{\omega}$, $[\underline{S}_{\alpha_j}^{\omega},\overline{S}_{\alpha_j}^{\omega}]$, which can bound the value of $\langle S_{\alpha_j}^{\omega}\rangle$ with a probability of at least $1-\epsilon$. Explicitly, we have $\delta_{\alpha_j}^{\omega}=\delta(N_{\alpha_j}^{\omega}S_{\alpha_j}^{\omega}, \epsilon)$ with the function $\delta(x,y)=[-\ln(y/2)+\sqrt{(\ln(y/2))^2-8 \ln(y/2) x}]/(2x)$. With the mean values $\langle s_1^{\omega,L}\rangle$ defined in Eq.(\ref{eq:Ls1mean}), the lower bounds of $s_{1,\alpha_2}^{\alpha}(\alpha=Z,X)$ can be calculated with
\begin{equation}\label{eq:Ls1}
  s_{1,\alpha_2}^{\alpha,L}=\langle s_1^{\alpha,L}\rangle (1-\delta_{1,\alpha_2}),
\end{equation}
where $\delta_{1,\alpha_2}=\delta(N_{1,\alpha_2}^{\alpha}\langle S_{1}^{\alpha,L}\rangle, \epsilon)$. Here and after, we define $N_{k,\alpha_j}^{\omega}=a_{k,\alpha_j} p_{\alpha_j} q^{\omega} N_t$ as the number of $k$-photon pulses prepared in source $\alpha_j$ and measured in the basis $\omega$.

Second, we can also formulate the phase-flip error rate of single-photon states. Explicitly, we have
\begin{equation}\label{eq:Ue1}
  e_{1,Z}^{ph}=e_{1,X_1}^{X}\leq e_{1,X_1}^{X,U}=\frac{\overline{T}_{X_1}^{X} -a_{0,X_1} \langle s_{0}^{X}\rangle (1-\delta_{0,X_1}^{X})/2}{a_{1,X_1} s_{1,X_1}^{X,L}},
\end{equation}
where $\delta_{0,X_1}^{X}=\delta(N_{0,X_1}^{X}\langle s_0^{X}\rangle ,\epsilon)$. In a finite-key-size case, we can apply the large data size approximation of the random sampling method to upper bound the phase error rate $e_1^{p,Z}$ of single-photon pulses prepared and measured in the $Z$ basis with the failure probability $\epsilon$
\begin{equation}\label{eq:Ue1p}
  e_1^{p,Z}\leq e_1^{p,Z,U}=e_{1,X_1}^{X,U} +\theta_Z^{X},
\end{equation}
where $\theta_Z^{X}=\sqrt{n_{\theta}/d_{\theta}}$ with $d_{\theta}=\frac{(1-g_X)g_X \ln 2}{2(1-e_1)e_1}$, $n_{\theta}=-\log[\epsilon \sqrt{e_1 (1-e_1)n_X n_Z/(n_X+n_Z)}]/(n_X+n_Z)$, and $g_X=\frac{n_X}{n_X+n_Z}$. Here we write $n_X=N_{1,X_1}^{X}, n_Z=N_{1,Z_2}^{Z}$ and $e_1=e_{1,X_1}^{X,U}$ for simplicity. Note that $e_1^{p,Z,U}$ is a function of $\langle s_0^X\rangle$. Straightly, we can also formulate the upper bound of the phase-flip error rate of single-photon counts in the $X$ basis, being denoted by $e_{1}^{p,X,U}$. We omit the explicit formula here since it is just trivially written analogically to Eq.(\ref{eq:Ue1p}).

Note that $\langle s_0^X\rangle$ (or $\langle s_0^Z\rangle$) is the common variable in both quantities $s_{1,X_2}^{X,L}$ and $e_1^{p,Z,U}$ ( or quantities $s_{1,Z_2}^{Z,L}$ and $e_1^{p,X,U}$) shown in Eq.(\ref{eq:Ls1}) and Eq.(\ref{eq:Ue1p}) respectively. We need to know the range of them for the final key rate calculation. In the 4-intensity protocol without the assumption of vacuum, we can lower bound $\langle s_0^{\omega}\rangle$ by
\begin{equation}\label{eq:Ls0Xmeanf}
  \langle s_0^{\omega}\rangle \geq \langle s_0^{\omega,L}\rangle =\max_{\alpha=X,Z}\{ \langle s_0^{\omega,L}\rangle(\alpha),0\},
\end{equation}
where
\begin{equation}\label{eq:Ls0Xmean}
  \langle s_0^{\omega}\rangle \geq \langle s_0^{\omega,L}\rangle(\alpha)=\frac{a_{1,\alpha_2} \underline{S}_{\alpha_1}^{\omega} -a_{1,\alpha_1} \overline{S}_{\alpha_2}^{\omega}}{A_{\alpha_1 \alpha_2}^{0,1}},
\end{equation}
and $A_{\alpha_1 \alpha_2}^{0,1}=a_{0,\alpha_1}a_{1,\alpha_2}-a_{0,\alpha_2}a_{1,\alpha_1}$. By simply attributing all the errors to the vacuum pulses, we can upper bound of $\langle s_0^X\rangle$ with
\begin{equation}\label{eq:Us0Xmeanf}
  \langle s_0^{\omega}\rangle \leq \langle s_0^{\omega,U}\rangle =\min \{ 2\overline{T}_{\omega_1}^{\omega}/a_{0,\omega_1}, \overline{S}_{Z_1}^{\omega}/a_{0,Z_1}, \overline{S}_{X_1}^{\omega}/a_{0,X_1}\}.
\end{equation}

With these preparations, the final key rate of the 4-intensity protocol can be calculated with the following worst-case estimation
\begin{equation}\label{eq:FKeyRate}
  R=\min_{\langle s_0^{Z}\rangle,\langle s_0^{X}\rangle}[ \mathcal{R}(\langle s_0^{Z}\rangle,\langle s_0^{X}\rangle)]
\end{equation}
over the region for all possible values of $\langle s_0^{Z}\rangle$ and $\langle s_0^{X}\rangle$ in $[\langle s_0^{Z,L}\rangle,\langle s_0^{Z,U}\rangle]$ and $[\langle s_0^{X,L}\rangle,\langle s_0^{X,U}\rangle]$, respectively. Here
\begin{equation}\label{eq:KeyRateFun}
  \mathcal{R}(\langle s_0^{Z}\rangle,\langle s_0^{X}\rangle)=\mathcal{R}_Z(\langle s_0^{Z}\rangle,\langle s_0^{X}\rangle) +\mathcal{R}_X(\langle s_0^{Z}\rangle,\langle s_0^{X}\rangle),
\end{equation}
and
\begin{equation}\label{eq:KeyRateZX}
  \mathcal{R}_{\alpha}(\langle s_0^{Z}\rangle,\langle s_0^{X}\rangle)= p_{\alpha_2}q^{\alpha} \{ a_{1,\alpha_2} s_{1,\alpha_2}^{\alpha,L} [1-H(e_{1}^{p,\alpha,U})] -f S_{\alpha_2}^{\alpha} H(E_{\alpha_2}^{\alpha})\},
\end{equation}
for $\alpha=Z,X$. Here $f$ is the efficiency factor of the error-correction method used, $H(x)=-x\log_2(x)-(1-x)\log_2(1-x)$ is the binary Shannon entropy function. Note that in such a case we need to calculate the final key rate with two variables $\langle s_0^{Z}\rangle$ and $\langle s_0^{X}\rangle$ jointly.

\section*{Experiment}
\begin{figure*}[hbtp]
	\centering
        \includegraphics[width=1\textwidth]{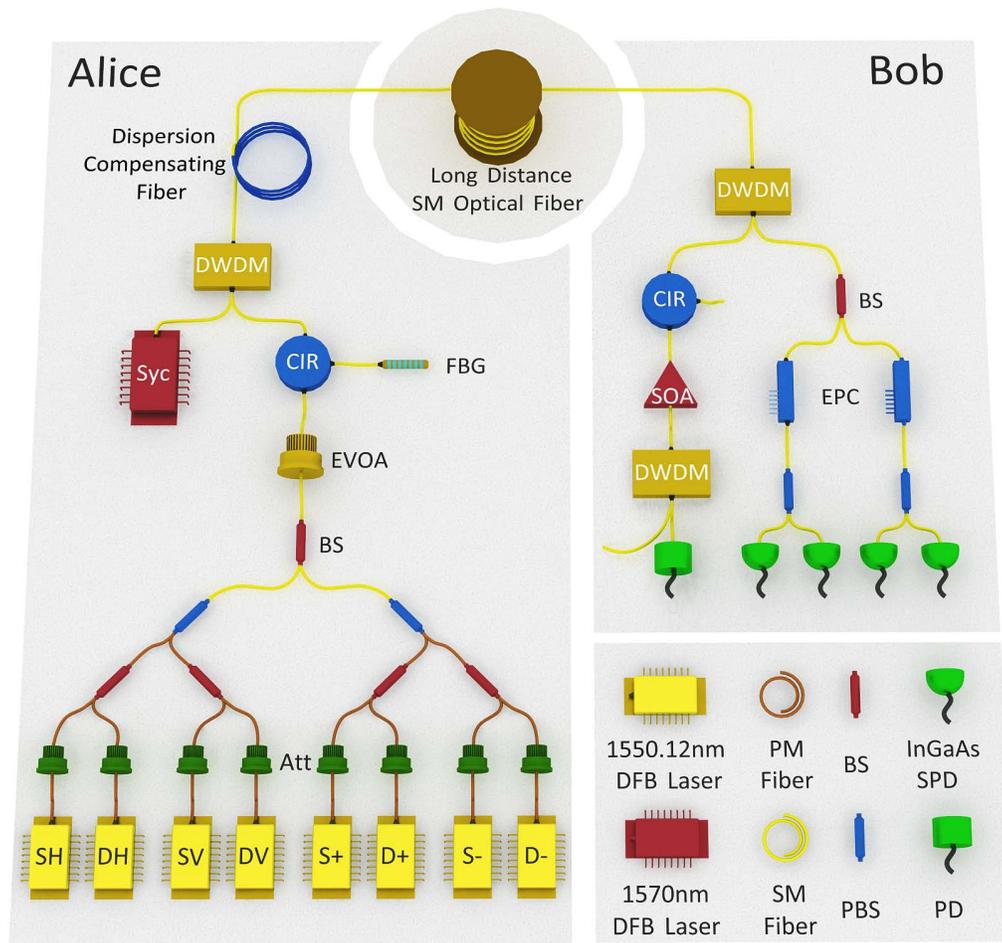}
\caption{Schematic layout of the experiment. DFB Laser: distributed feedback laser, Att: manual attenuator, PM: polarization maintaining, SM: single mode, BS: beam splitter, PBS: polarization beam splitter, EVOA: electrical variable optical attenuator, FBG: fiber Bragg grating, DCF: dispersion compensating fiber, DWDM: dense wavelength division multiplexer, SOA: semiconductor optical amplifier, EPC: electric polarization controllers, PD: photoelectric detector, APD: avalanche photodiode. }
	\label{fig:setup}
\end{figure*}

The polarization encoding is implemented in our experiment. FIG.\ref{fig:setup} illustrates the scheme of our experimental setup. The Z and X basis consists of ${\{|H\rangle, |V\rangle\}}$ and ${\{|+\rangle,|-\rangle\}}$, respectively, as four states for the standard BB84 protocol. The signals are generated at a system clock rate of 625 MHz by 8 DFB lasers, half of which are used for generating signal state and the rest are used for decoy state. Alice encodes her qubits in Z or X basis in accordance with random bit values generated beforehand. The pulse width is about 100~ps and its wavelength center is at 1550.12~nm. These pulses are naturally phase randomized due to direct modulation onto DFB lasers. Utilizing 8 manual attenuators after each DFB laser, Alice realizes the intensity ratio of two intensities in each basis approximately. None of the DFB laser generated the pulse when vacuum pulse are need. Four PMBSs, two PMPBSs and a SMBS server for guiding pulses from different diodes to one optical fiber. The optical pulse intensity is strongly attenuated to single-photon level via an EVOA.

A 10~GHz FBG is inserted at Alice for three reasons. First, it guarantees that the spectrum of 8 DFB lasers are overlapped in a narrow range to get rid of the loopholes exploiting the pulses wavelength discrepancies. The second issue is that it achieves fine adjustment of the state intensities coordinated with the precise temperature control of the DFB laser. At last, it reduces the chromatic dispersion effects in long single-mode optical fiber. A suitable Dispersion Compensating Fiber is installed to compensate dispersion effects further and compress the pulses width, which guarantees that the pulse width is smaller than the detector effective gate width after long distance propagation.

The synchronization pulses are generated by a 1570~nm DFB laser operating at 100~kHz. In order to synchronize the entire experimental systems and reduce optical fiber costs, the synchronization pulses emitted from Alice are multiplexed with signal pulses by a 100~G DWDM and transmitted through the same single-mode optical fiber  to Bob. A SOA is utilized to amplify the intensity of synchronization pulses to guarantee that Bob's PD receive sufficient optical power. A DWDM inserted before his PD is typically introduced filter undesired noise from the SOA.

Naively, Bob passively selects the measurement basis by a 1$\times$2 SMBS with the splitting ratio of ${q_X}$. It indicates that the received photons are measured either on the X or Z basis randomly with probabilities of ${q_X}$ and ${q_Z} =1-{q_X}$, respectively. Cooled to $-50\,^{\circ}\mathrm{C}$, four InGaAs APDs operating in gated Geiger mode are used to detect signals at 1.25~GHz gating frequency. The effective gating window width is 180~ps and the dead time is 500~ps, which is an optimal trade-off between the detection efficiency and the after pulses rate. The detection efficiency is about 10\% at a dark count probability of $2.50\times10^{-7}$ per gate. For convenience, we inserted two 3~dB or even 10~dB attenuations, one before each of two APDs for X basis, to get a larger efficiency asymmetry and demonstrate the effectiveness of difference protocols. We thus regard the attenuations as a part of the APDs.

Alice and Bob have to develop a stable polarization reference frame initially owing to the polarization mode dispersion (PMD) effects in long distance single-mode optical fiber. Bob applies corresonding DC voltage on a pair of EPCs to align Alice's polarization states to the polarizing axes of the PBSs inserted before the APDs. The optical misalignment error rate $e_d$ is around 1.5\%. Note that the optical misalignment error rate of Z and X basis are independent. The polarization can remain stable for more than 20 minutes, which is long enough for our experiment.

\section*{Results}
Using same system parameters in Table\ref{tab:para} to perform a numerical optimization for consistency and taking the effects of statistical fluctuations into account, we implement three decoy-state BB84 protocols: (I) traditional 3-intensity protocol  \cite{wang2007quantum} with basis detector efficiency symmetry, where Bob reduce higher detecotor efficiency to balance asymemetry $\eta_Z = \eta_X$; (II) 3-intensity protocol with basis detector efficiencies asymmetry \cite{yu2016reexamination}, where in both bases, Alice select the same intensities and proportions, and Bob measures the received pulses with the same probabilities, that is, ${q_Z} = {q_X} = 50\%$; (III) 4-intensity protocol\cite{yu2016reexamination}, where ${q_Z} \neq {q_X}$. In all protocols, the signal pulses  $\mu_{Z_2}$ and $\mu_{X_2}$ are used for key generation, while other intensity pulses are used as decoy states to estimate the amount of privacy amplification necessary. The extra insertion loss in Bob is about 2.5-2.7~dB due to different BSs in different protocols. Thanks to the high clock rate, sufficient signal pulses are send by Alice during an uninterrupted session lasting 16.16 s to calculate the final key rate. We repeat the experiment 30 times and calculate the average and variance of the final key rate, which is shown in FIG.\ref{fig:res1}. Details of the  main implementation parameters and results are shown in the Supplemental Material.

In the first experiment, we set the detector efficiency of the InGaAs APD $\eta_Z = 10\%$ and $\eta_X = 5\%$, that is, the asymmetry $\eta_Z/\eta_X = 2$, and change the distance between Alice and Bob from 87~km to 150~km. The results are shown in FIG.\ref{fig:res1}~(a). Consequently, 4-intensity protocol dramatically gives measurable advantage over two types of 3-intensity protocol. For example, 4-intensity protocol obtain a key rate of 39~kbps in 87~km, which is 3.0 times that of 3-intensity protocol and 4.8 times that of 3-intensity protocol with basis detector efficiencies symmetry. And 4-intensity yield a secret key rate of 36.7~bps in a maximal distance of 150~km. In contrast, not even a bit of secure key can be extracted with both two types of 3-intensity. In the second experiment, we increase the mismatch on purpose and set $\eta_Z = 10\%$ and $\eta_X = 1\%$. FIG. \ref{fig:res1}(b) presents the experiment results. The experiment data of the 87~km case is used as an example to demonstrate the improvement of 4-intensity protocol. 4-intensity protocol obtain a key rate of 20~kbps in 87~km, which is 3.6 times that of 3-intensity protocol and 33.2 times that of 3-intensity protocol with basis detector efficiencies symmetry. 

\section*{Conclusion}
In summary, we have demonstrated, for the first time, an implementation of decoy-state QKD system with asymmetric basis detector efficiencies by the recent 4-intensity decoy-state optimization protocol. The secure key rate is higher than previous traditional 3-intensity protocols with unbiased bases results by 1.9 to 33.2 times. Besides, our results ruled out an implicitly assumption in QKD that the efficiency of Z basis and X basis are restricted to be same. Therefore, the implementation is an excellent candidate for future quantum key distribution.
\section*{Acknowledgments}
  This work was supported by the National Key R\&D Program of China (2017YFA0303903), the National Natural Science Foundation of China (Grant No. 61875182), and Anhui Initiative in Quantum Information Technologies and Fundamental Research Funds for the Central Universities (WK2340000083).
\bibliography{Redecoyref}

\renewcommand\subsection[1]{
\vspace{\baselineskip}
\textbf{#1}
\vspace{0.5\baselineskip}
}

\clearpage

\clearpage

\begin{figure}[hbtp]
	\centering
	\includegraphics[width=0.7\textwidth]{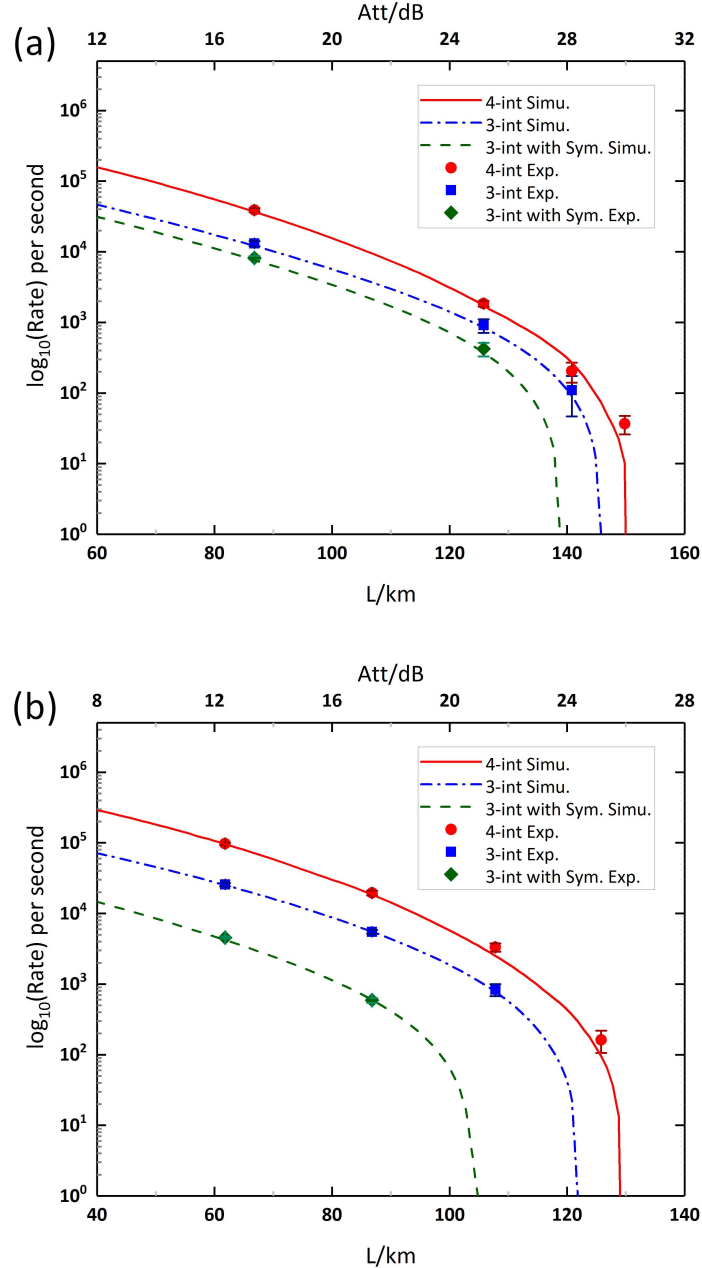}
	\caption{Experimentally (symbols) and simulated (solid lines) secret key rates in bps versus the transmission distance in standard optical fiber.  (a) the detector efficiency \{$\eta_Z, \eta_X$\} is fixed at \{10\%, 5\%\},while the experimental transmission distance are selected at 87, 126, 141 and 150~km. (b) the detector efficiency \{$\eta_Z, \eta_X$\} is fixed at \{10\%, 1\%\}, while the experimental transmission distance are selected at 62, 87, 107 and 126~km. Blue squares, green diamonds and red circles, respectively, refers to (I) traditional 3-intensity protocol \cite{wang2007quantum}  basis detector efficiencies symmetry, where Bob reduce higher detecotor effiency to balance asymemetry $\eta_Z = \eta_X = 5\%~~or~~1\%$; (II) 3-intensity protocol with basis detector efficiencies asymmetry \cite{yu2016reexamination}, where $u_{X_2} = u_{Z_2}, u_{X_1} = u_{Z_1}, {q_z} = {q_x}, \eta_Z \neq \eta_X$;  (III) 4-intensity protocol\cite{yu2016reexamination}, where $u_{X_2} \neq u_{Z_2}, u_{X_1} \neq u_{Z_1}, {q_Z} \neq {q_X}, \eta_Z \neq \eta_X$. The experimental results are the average and variance (1 standard deviation, assuming Poissonian detection statistics) of the final key rate calculated by 30 experiments. The advantage of the 4-intensity protocol is clearly verified by the experimental results, especially in a large degree of basis detector efficiencies asymmetry. The results also confirm the excellent stability of three protocols used here. }
	\label{fig:res1}
\end{figure}
\clearpage

\begin{table}[h]
\caption{List of parameters characterized for numerical optimization: detector dark count rate $s_0$,detector After Pulses rate $Ar$, Detector dead time $t_d$ in second, misalignment-error probability $e_d$, channel loss coefficient $\alpha$ in dB/km, error-correction efficiency $f$, security parameter $\epsilon$, and the total number of laser pulses $N$  }

\begin{tabular}{p{2cm}<{\centering}p{1cm}<{\centering}p{2cm}<{\centering}p{1cm}<{\centering}p{1cm}<{\centering}p{1cm}<{\centering}p{1cm}<{\centering}p{1cm}<{\centering}}
\hline\hline

\specialrule{0em}{1.5pt}{1.5pt}
   $s_0$ &     $Ar$ &       $t_d$ &   $e_d$ &    $\alpha$ &     $f$ &    $\epsilon$ &     $N$ \\
\specialrule{0em}{1.5pt}{1.5pt}
\hline
\specialrule{0em}{1.2pt}{1.2pt}
  $2.50\times10^{-7}$ &       $1\%$ &     $5.00\times10^{-7}$ &        $1.5\%$ &       $0.2$ &       $1.14$ &   $10^{-10}$ &   $10^{10}$ \\
\specialrule{0em}{1.2pt}{1.2pt}
\hline\hline
\end{tabular}
\label{tab:para}
\end{table}

\onecolumngrid
\begin{center}[!ht]
\begin{sidewaystable}
\centering
\caption{List of the main implementation parameters. Here, The notation $u_{\alpha_j}$ shown in the first column denotes the intensity of the coherent source $\alpha_j$. $p_{\alpha_j}$ is the probability to use the source $\alpha_j$ in the protocol. $p_0$ is the probability to choose the vacuum source. $q_X$ is the probability that Bob measures the received pulses in the X basis.}
\begin{tabular}{p{2cm}<{\centering}|p{2cm}<{\centering}p{2cm}<{\centering}p{2cm}<{\centering}|p{2cm}<{\centering}p{2cm}<{\centering}p{2cm}<{\centering}|p{2cm}<{\centering}p{2cm}<{\centering}|p{3cm}<{\centering}}

\hline
\hline
\multicolumn{1}{c|}{Parameters} &  \multicolumn{3}{c|}{62km(12.4dB)} &  \multicolumn{3}{c|}{87km(17.4dB)} & \multicolumn{2}{c|}{107km(21.4dB)} & 126km(25.2dB) \\
\cline{1-10}
\multicolumn{1}{c|}{10\%1\%} & 3INT(1\%1\%) &       3INT &       4INT & 3INT(1\%1\%) &       3INT &       4INT &       3INT &       4INT &       4INT \\
\hline
      $u_{X_2}$ &     0.134  &     0.145  &     0.418  &     0.515  &     0.501  &     0.714  &     0.499  &     0.671  &     0.560  \\
      $u_{Z_2}$ &     0.134  &     0.145  &     0.470  &     0.515  &     0.501  &     0.453  &     0.499  &     0.449  &     0.429  \\
      $u_{X_1}$ &     0.522  &     0.492  &     0.175  &     0.148  &     0.169  &     0.206  &     0.179  &     0.206  &     0.204  \\
      $u_{Z_1}$ &     0.522  &     0.492  &     0.026  &     0.148  &     0.169  &     0.041  &     0.179  &     0.054  &     0.065  \\
      $p_{X_2}$ &     0.406  &     0.418  &     0.030  &     0.327  &     0.365  &     0.032  &     0.272  &     0.040  &     0.034  \\
      $p_{Z_2}$ &     0.406  &     0.418  &     0.754  &     0.327  &     0.365  &     0.693  &     0.272  &     0.577  &     0.288  \\
      $p_{X_1}$ &     0.083  &     0.064  &     0.110  &     0.154  &     0.107  &     0.132  &     0.184  &     0.170  &     0.291  \\
      $p_{Z_1}$ &     0.083  &     0.064  &     0.105  &     0.154  &     0.107  &     0.143  &     0.184  &     0.214  &     0.387  \\
      $p_{0}$ &     0.022  &     0.036  &     NULL  &     0.038  &     0.055  &     NULL  &     0.088  &     NULL  &     NULL  \\
       $q_X$ &       50\% &       50\% &       15\% &       50\% &       50\% &       24\% &       50\% &       36\% &       55\% \\

\hline
\multicolumn{1}{c|}{10\%5\%} &  \multicolumn{3}{c|}{87km(17.4dB)} &  \multicolumn{3}{c|}{126km(25.2dB)} & \multicolumn{2}{c|}{141km(28.2dB)} & 150km(30.0dB) \\
\hline
      $u_{X2}$ &     0.524  &     0.521  &     0.473  &     0.513  &     0.513  &     0.451  &     0.512  &     0.621  &     0.572  \\
      $u_{Z2}$ &     0.524  &     0.521  &     0.516  &     0.513  &     0.513  &     0.445  &     0.512  &     0.459  &     0.523  \\
      $u_{X1}$ &     0.127  &     0.125  &     0.170  &     0.149  &     0.153  &     0.196  &     0.151  &     0.312  &     0.262  \\
      $u_{Z1}$ &     0.127  &     0.125  &     0.020  &     0.149  &     0.153  &     0.041  &     0.151  &     0.070  &     0.111  \\
      $p_{X2}$ &     0.421  &     0.428  &     0.019  &     0.294  &     0.331  &     0.037  &     0.200  &     0.055  &     0.098  \\
      $p_{Z2}$ &     0.421  &     0.428  &     0.772  &     0.294  &     0.331  &     0.531  &     0.200  &     0.338  &     0.175  \\
      $p_{X1}$ &     0.069  &     0.062  &     0.116  &     0.183  &     0.149  &     0.172  &     0.262  &     0.235  &     0.228  \\
      $p_{Z1}$ &     0.069  &     0.062  &     0.094  &     0.183  &     0.149  &     0.259  &     0.262  &     0.372  &     0.499  \\
       $p_0$ &     0.020  &     0.020  &     NULL  &     0.045  &     0.041  &     NULL  &     0.076  &     NULL  &     NULL  \\
       $q_X$ &       50\% &       50\% &       13\% &       50\% &       50\% &       24\% &       50\% &       34\% &       50\% \\
\hline
\hline
\end{tabular}
\end{sidewaystable}
\end{center}

\onecolumngrid
\begin{center}[!ht]
\begin{sidewaystable}
\centering
\caption{List of the  average and variance (1 standard deviation, assuming Poissonian detection statistics) of the total gains and error gains in the cases of \{$\eta_z=10\%, \eta_x$=5\%\}. The notation ${\alpha_j}\beta$ shown in the second column denotes the pulse from Alice source ${\alpha_j}$ and the basis $\beta$ chosen by Bob, respectively.}

\newcommand{\tabincell}[2]{\begin{tabular}{@{}#1@{}}#2\end{tabular}}
\begin{tabular}{p{1cm}<{\centering}p{1cm}<{\centering}|p{2cm}<{\centering}p{2cm}<{\centering}p{2cm}<{\centering}|p{2cm}<{\centering}p{2cm}<{\centering}p{2cm}<{\centering}|p{2cm}<{\centering}p{2cm}<{\centering}|p{2cm}<{\centering}}
\hline
\hline
\multicolumn{2}{c|}{\multirow{2}{2cm}{Parameters\\$\quad$10\%5\%}} &                                          \multicolumn{3}{c|}{87km(17.4dB)} &                                          \multicolumn{3}{c|}{126km(25.2dB)} &               \multicolumn{2}{c|}{141km(28.2dB)} & \multicolumn{1}{c}{150km(30.0dB)} \\
\cline{3-11}
 \multicolumn{2}{c}{} & \multicolumn{1}{|c}{3INT(5\%5\%)} & \multicolumn{1}{c}{3INT} & \multicolumn{1}{c}{4INT} & \multicolumn{1}{|c}{3INT(5\%5\%)} & \multicolumn{1}{c}{3INT} & \multicolumn{1}{c}{4INT} & \multicolumn{1}{|c}{3INT} & \multicolumn{1}{c}{4INT} & \multicolumn{1}{|c}{4INT} \\
\hline
\multicolumn{1}{c|}{\multirow{12}{1cm}{Total\\Gains}} &        $X_2X$ &  \tabincell{c}{551418.8\\$\pm$13513.4}  &  \tabincell{c}{556153.6\\$\pm$3713.5}  &    \tabincell{c}{5709.1\\$\pm$177.0}  &   \tabincell{c}{64974.3\\$\pm$499.2}  &   \tabincell{c}{73655.5\\$\pm$1053.5}  &    \tabincell{c}{3698.1\\$\pm$76.9}  &   \tabincell{c}{22636.7\\$\pm$202.6}  &    \tabincell{c}{5335.4\\$\pm$103.4}  &    \tabincell{c}{8441.0\\$\pm$156.3}  \\

\multicolumn{ 1}{c|}{} &        $X_1X$ &   \tabincell{c}{22246.4\\$\pm$723.4}  &   \tabincell{c}{19234.5\\$\pm$206.0}  &   \tabincell{c}{13826.9\\$\pm$251.1}  &   \tabincell{c}{12376.7\\$\pm$323.4}  &   \tabincell{c}{10122.4\\$\pm$425.9}  &    \tabincell{c}{7499.0\\$\pm$302.5}  &    \tabincell{c}{9422.2\\$\pm$130.1} &   \tabincell{c}{11369.8\\$\pm$302.5}  &    \tabincell{c}{9416.2\\$\pm$94.4}  \\

\multicolumn{ 1}{c|}{} &        $X_0X$ &     145.6$\pm$18.8  &     143.6$\pm$10.8  &       NULL  &     144.2$\pm$12.9  &     130.7$\pm$14.3  &       NULL  &     210.2$\pm$16.5  &       NULL  &       NULL  \\

\multicolumn{ 1}{c|}{} &        $Z_2Z$ &  \tabincell{c}{532257.2\\$\pm$13606.8}  & \tabincell{c}{1072606.3\\$\pm$6749.1}  & \tabincell{c}{3340950.6\\$\pm$27645.2}  &   \tabincell{c}{63247.3\\$\pm$970.7}  &  \tabincell{c}{143018.1\\$\pm$3543.1}  &  \tabincell{c}{302314.4\\$\pm$2091.8}  &   \tabincell{c}{43933.5\\$\pm$613.3}  &   \tabincell{c}{86861.5\\$\pm$1534.9}  &   \tabincell{c}{25800.8\\$\pm$285.3}  \\

\multicolumn{ 1}{c|}{} &        $Z_1Z$ &   \tabincell{c}{22075.4\\$\pm$648.5}  &   \tabincell{c}{38122.8\\$\pm$348.8}  &   \tabincell{c}{18908.5\\$\pm$248.9}  &   \tabincell{c}{12376.7\\$\pm$177.4}  &   \tabincell{c}{19893.0\\$\pm$314.7}  &   \tabincell{c}{15815.0\\$\pm$382.9}  &   \tabincell{c}{18067.7\\$\pm$379.3}  &   \tabincell{c}{16306.8\\$\pm$277.3}  &   \tabincell{c}{17446.6\\$\pm$218.3}  \\

\multicolumn{ 1}{c|}{} &        $Z_0Z$ &     144.3$\pm$17.5  &     264.4$\pm$19.3  &       NULL  &     140.1$\pm$12.5  &     155.9$\pm$14.9  &       NULL  &     212.3$\pm$17.2  &       NULL  &       NULL  \\

\multicolumn{ 1}{c|}{} &        $X_2Z$ &  \tabincell{c}{547810.9\\$\pm$12536.7}  & \tabincell{c}{1068918.4\\$\pm$9105.1}  &   \tabincell{c}{74118.7\\$\pm$835.2}  &   \tabincell{c}{65447.0\\$\pm$519.5}  &  \tabincell{c}{145119.9\\$\pm$1851.1}  &   \tabincell{c}{21745.1\\$\pm$271.7}  &   \tabincell{c}{44756.9\\$\pm$736.4}  &   \tabincell{c}{19220.3\\$\pm$293.2}  &   \tabincell{c}{15890.5\\$\pm$164.8}  \\

\multicolumn{ 1}{c|}{} &        $X_1Z$ &   \tabincell{c}{21825.1\\$\pm$708.7}  &   \tabincell{c}{38409.7\\$\pm$298.8}  &  \tabincell{c}{168526.6\\$\pm$3115.7}  &   \tabincell{c}{12051.5\\$\pm$301.1}  &   \tabincell{c}{19636.4\\$\pm$585.8}  &   \tabincell{c}{43211.6\\$\pm$1442.8}  &   \tabincell{c}{17560.9\\$\pm$336.7}  &   \tabincell{c}{39677.8\\$\pm$683.9}  &   \tabincell{c}{16696.5\\$\pm$252.0}  \\

\multicolumn{ 1}{c|}{} &        $X_0Z$ &     144.8$\pm$16.4  &     254.9$\pm$16.8  &       NULL  &     145.2$\pm$14.0  &     154.7$\pm$15.2  &       NULL  &     207.6$\pm$14.9  &       NULL  &       NULL  \\

\multicolumn{ 1}{c|}{} &        $Z_2X$ &  \tabincell{c}{539397.7\\$\pm$14626.5}  &  \tabincell{c}{538231.6\\$\pm$4594.5}  &  \tabincell{c}{264307.1\\$\pm$3417.4}  &   \tabincell{c}{64809.0\\$\pm$926.4}  &   \tabincell{c}{71796.5\\$\pm$2782.3}  &   \tabincell{c}{50918.5\\$\pm$515.7}  &   \tabincell{c}{22170.3\\$\pm$274.8}  &   \tabincell{c}{24155.9\\$\pm$424.6}  &   \tabincell{c}{13267.3\\$\pm$218.8}  \\

\multicolumn{ 1}{c|}{} &        $Z_1X$ &   \tabincell{c}{22203.6\\$\pm$538.0}  &   \tabincell{c}{19728.7\\$\pm$247.8}  &    \tabincell{c}{1843.7\\$\pm$49.6}  &   \tabincell{c}{12401.1\\$\pm$163.2}  &   \tabincell{c}{10292.6\\$\pm$236.7}  &    \tabincell{c}{3453.1\\$\pm$73.3}  &    \tabincell{c}{9551.0\\$\pm$126.3}  &    \tabincell{c}{5520.6\\$\pm$97.3}  &   \tabincell{c}{10023.2\\$\pm$100.0}  \\

\multicolumn{ 1}{c|}{} &        $Z_0X$ &     144.5$\pm$17.4  &     142.5$\pm$14.1  &       NULL  &     138.1$\pm$11.5  &     131.4$\pm$13.9  &       NULL  &     211.5$\pm$17.8  &       NULL  &       NULL  \\
\hline
\multicolumn{1}{c|}{\multirow{6}{1cm}{Error\\Gains}} &        $X_2X$ &    \tabincell{c}{9199.7\\$\pm$2236.8}  &    \tabincell{c}{8747.8\\$\pm$784.3}  &     \tabincell{c}{143.9\\$\pm$14.4}  &    \tabincell{c}{1664.3\\$\pm$147.5}  &    \tabincell{c}{1840.7\\$\pm$312.9}  &     \tabincell{c}{154.9\\$\pm$16.2}  &     \tabincell{c}{853.1\\$\pm$148.2}  &     \tabincell{c}{216.6\\$\pm$30.8}  &     \tabincell{c}{340.3\\$\pm$27.3} \\

\multicolumn{ 1}{c|}{} &        $X_1X$ &     741.8$\pm$129.3  &     635.9$\pm$36.2  &     600.8$\pm$30.1  &     706.8$\pm$32.1  &     584.2$\pm$62.7  &     554.0$\pm$38.6  &     841.0$\pm$92.5  &     748.8$\pm$66.2  &     634.1$\pm$29.1  \\

\multicolumn{ 1}{c|}{} &        $X_0X$ &      73.4$\pm$9.2  &      72.9$\pm$8.2  &       NULL  &      74.1$\pm$9.8  &      64.9$\pm$10.3  &       NULL  &     106.5$\pm$9.6  &       NULL &       NULL  \\

\multicolumn{ 1}{c|}{} &        $Z_2Z$ &   \tabincell{c}{12363.2\\$\pm$2150.4}  &   \tabincell{c}{22007.5\\$\pm$3341.7}  &   \tabincell{c}{62990.5\\$\pm$4278.1}  &    \tabincell{c}{1925.5\\$\pm$289.2}  &    \tabincell{c}{3644.1\\$\pm$263.3}  &    \tabincell{c}{6533.8\\$\pm$640.6}  &    \tabincell{c}{1171.3\\$\pm$80.9}  &    \tabincell{c}{2410.9\\$\pm$204.2}  &     \tabincell{c}{797.1\\$\pm$62.7}  \\

\multicolumn{ 1}{c|}{} &        $Z_1Z$ &     \tabincell{c}{864.0\\$\pm$107.4}  &    \tabincell{c}{1361.4\\$\pm$111.4}  &    \tabincell{c}{1999.8\\$\pm$96.5}  &     \tabincell{c}{747.4\\$\pm$56.6}  &     \tabincell{c}{886.1\\$\pm$43.8}  &    \tabincell{c}{1357.3\\$\pm$75.9}  &     \tabincell{c}{974.3\\$\pm$48.4}  &    \tabincell{c}{1379.0\\$\pm$36.7}  &    \tabincell{c}{1533.3\\$\pm$57.3}  \\

\multicolumn{ 1}{c|}{} &        $Z_0Z$ &      71.8$\pm$9.5  &     133.1$\pm$12.6  &       NULL  &      70.7$\pm$9.6  &      79.7$\pm$10.7  &       NULL  &     106.9$\pm$12.1  &       NULL  &       NULL  \\
\hline
\multicolumn{ 2}{c|}{$R$} &   \tabincell{c}{1.32E-05\\$\pm$8.07E-07} &   \tabincell{c}{2.12E-05\\$\pm$1.77E-06} &   \tabincell{c}{6.30E-05\\$\pm$4.02E-06} &   \tabincell{c}{6.81E-07\\$\pm$1.50E-07} &   \tabincell{c}{1.48E-06\\$\pm$3.25E-07} &   \tabincell{c}{2.98E-06\\$\pm$2.77E-07} &   \tabincell{c}{1.78E-07\\$\pm$1.03E-07} &   \tabincell{c}{3.30E-07\\$\pm$1.04E-07} &   \tabincell{c}{5.93E-08\\$\pm$1.71E-08} \\
\hline

\hline
\end{tabular}
\end{sidewaystable}
\end{center}

\onecolumngrid
\begin{center}[!ht]
\begin{sidewaystable}
\centering
\caption{List of the average and variance (1 standard deviation, assuming Poissonian detection statistics) of the total gains and error gains in the cases of \{$\eta_z=10\%, \eta_x$=1\%\}. The notation ${\alpha_j}\beta$ shown in the second column denotes the pulse from Alice source ${\alpha_j}$ and the basis $\beta$ chosen by Bob, respectively.}

\newcommand{\tabincell}[2]{\begin{tabular}{@{}#1@{}}#2\end{tabular}}
\begin{tabular}{p{1cm}<{\centering}p{1cm}<{\centering}|p{2cm}<{\centering}p{2cm}<{\centering}p{2cm}<{\centering}|p{2cm}<{\centering}p{2cm}<{\centering}p{2cm}<{\centering}|p{2cm}<{\centering}p{2cm}<{\centering}|p{2cm}<{\centering}}
\hline
\hline
\multicolumn{2}{c|}{\multirow{2}{2cm}{Parameters\\$\quad$10\%1\%}} &                                          \multicolumn{3}{c|}{62km(12.4dB)} &                                          \multicolumn{3}{c|}{87km(17.4dB)} &               \multicolumn{2}{c|}{107km(21.4dB)} & \multicolumn{1}{c}{126km(25.2dB)} \\
\cline{3-11}
 \multicolumn{ 2}{c}{} & \multicolumn{1}{|c}{3INT(1\%1\%)} & \multicolumn{1}{c}{3INT} & \multicolumn{1}{c}{4INT} & \multicolumn{1}{|c}{3INT(1\%1\%)} & \multicolumn{1}{c}{3INT} & \multicolumn{1}{c}{4INT} & \multicolumn{1}{|c}{3INT} & \multicolumn{1}{c}{4INT} & \multicolumn{1}{|c}{4INT} \\
\hline
\multicolumn{1}{c|}{\multirow{12}{1cm}{Total\\Gains}} &       $X_2X$ &  \tabincell{c}{330399.5\\$\pm$6573.9}  &  \tabincell{c}{312310.7\\$\pm$5027.8}  &    \tabincell{c}{6007.4\\$\pm$134.4}  &   \tabincell{c}{83854.6\\$\pm$1022.9}  &   \tabincell{c}{93979.9\\$\pm$1661.1}  &    \tabincell{c}{5873.1\\$\pm$160.5}  &   \tabincell{c}{28473.8\\$\pm$734.5}  &    \tabincell{c}{4037.7\\$\pm$81.4}  &    \tabincell{c}{1972.3\\$\pm$51.8}  \\
\multicolumn{1}{c|}{} &       $X_1X$ &   \tabincell{c}{18234.3\\$\pm$238.5}  &   \tabincell{c}{15305.9\\$\pm$174.6}  &    \tabincell{c}{9738.6\\$\pm$146.1}  &   \tabincell{c}{11901.2\\$\pm$307.5}  &    \tabincell{c}{9710.8\\$\pm$156.2}  &    \tabincell{c}{7366.3\\$\pm$99.2}  &    \tabincell{c}{7557.3\\$\pm$201.2}  &    \tabincell{c}{5715.3\\$\pm$149.5}  &    \tabincell{c}{6658.3\\$\pm$151.7}  \\
\multicolumn{1}{c|}{} &       $X_0X$ &     120.1$\pm$11.8  &     201.9$\pm$15.5  &       NULL  &     126.6$\pm$15.1  &     188.4$\pm$16.5  &       NULL  &     236.3$\pm$13.8  &       NULL  &       NULL  \\
\multicolumn{1}{c|}{} &       $Z_2Z$ &  \tabincell{c}{341589.9\\$\pm$5789.6}  & \tabincell{c}{2885578.7\\$\pm$26291.5}  & \tabincell{c}{8152239.7\\$\pm$153051.7}  &   \tabincell{c}{86854.9\\$\pm$1458.3}  &  \tabincell{c}{893901.8\\$\pm$12110.0}  & \tabincell{c}{2308003.1\\$\pm$41297.7}  &  \tabincell{c}{271688.1\\$\pm$3812.4}  &  \tabincell{c}{647788.3\\$\pm$14340.1}  &   \tabincell{c}{93277.2\\$\pm$1145.7}  \\
\multicolumn{1}{c|}{} &       $Z_1Z$ &   \tabincell{c}{18789.9\\$\pm$235.3}  &  \tabincell{c}{133707.8\\$\pm$435.9}  &   \tabincell{c}{70249.0\\$\pm$1136.4}  &   \tabincell{c}{12342.7\\$\pm$250.8}  &   \tabincell{c}{88989.3\\$\pm$1698.2}  &   \tabincell{c}{46308.4\\$\pm$270.7}  &   \tabincell{c}{65140.2\\$\pm$995.3}  &   \tabincell{c}{31450.7\\$\pm$607.3}  &   \tabincell{c}{21229.4\\$\pm$281.8}  \\
\multicolumn{1}{c|}{} &       $Z_0Z$ &     119.4$\pm$13.1  &    1027.6$\pm$48.4  &       NULL  &     127.0$\pm$13.7  &     589.6$\pm$35.6  &       NULL  &     456.8$\pm$18.6  &       NULL  &       NULL  \\
\multicolumn{1}{c|}{} &       $X_2Z$ &  \tabincell{c}{347785.7\\$\pm$1595.7}  & \tabincell{c}{2993230.8\\$\pm$4815.9}  &  \tabincell{c}{291752.5\\$\pm$2585.1}  &   \tabincell{c}{86737.4\\$\pm$1638.3}  &  \tabincell{c}{891406.4\\$\pm$8560.7}  &  \tabincell{c}{169139.3\\$\pm$2999.7}  &  \tabincell{c}{270912.9\\$\pm$2959.6}  &   \tabincell{c}{67148.5\\$\pm$1593.0}  &   \tabincell{c}{14662.1\\$\pm$221.0}  \\
\multicolumn{1}{c|}{} &       $X_1Z$ &   \tabincell{c}{18474.7\\$\pm$401.0}  &  \tabincell{c}{131317.7\\$\pm$1233.1}  &  \tabincell{c}{449001.8\\$\pm$7198.0}  &   \tabincell{c}{12382.3\\$\pm$307.3}  &   \tabincell{c}{88134.9\\$\pm$1180.9}  &  \tabincell{c}{195019.6\\$\pm$2437.9}  &   \tabincell{c}{65305.2\\$\pm$765.7}  &   \tabincell{c}{85263.0\\$\pm$2731.9}  &   \tabincell{c}{44475.4\\$\pm$1011.4}  \\
\multicolumn{1}{c|}{} &       $X_0Z$ &     121.7$\pm$10.9  &    1015.8$\pm$45.3  &       NULL  &     124.5$\pm$12.0  &     591.6$\pm$29.7  &       NULL  &     454.7$\pm$23.5  &       NULL  &       NULL  \\
\multicolumn{1}{c|}{} &       $Z_2X$ &  \tabincell{c}{339096.4\\$\pm$3402.7}  &  \tabincell{c}{321268.0\\$\pm$2897.1}  &  \tabincell{c}{173910.1\\$\pm$1725.4}  &   \tabincell{c}{82948.8\\$\pm$528.8}  &   \tabincell{c}{91943.3\\$\pm$1347.1}  &   \tabincell{c}{80238.4\\$\pm$1168.3}  &   \tabincell{c}{27745.8\\$\pm$690.8}  &   \tabincell{c}{39802.4\\$\pm$771.1}  &   \tabincell{c}{12868.8\\$\pm$292.3}  \\
\multicolumn{1}{c|}{} &       $Z_1X$ &   \tabincell{c}{18127.1\\$\pm$241.5}  &   \tabincell{c}{14025.7\\$\pm$268.3}  &    \tabincell{c}{1913.4\\$\pm$52.7}  &   \tabincell{c}{11864.5\\$\pm$185.8}  &    \tabincell{c}{9535.2\\$\pm$158.6}  &    \tabincell{c}{2174.9\\$\pm$46.4}  &    \tabincell{c}{7327.3\\$\pm$182.7}  &    \tabincell{c}{2696.9\\$\pm$71.8}  &    \tabincell{c}{4157.0\\$\pm$95.1}  \\
\multicolumn{1}{c|}{} &       $Z_0X$ &     121.1$\pm$13.2  &     199.8$\pm$14.6  &       NULL  &     126.7$\pm$19.0  &     185.3$\pm$13.1  &       NULL  &     239.2$\pm$16.5  &       NULL  &      NULL  \\
\hline
\multicolumn{1}{c|}{\multirow{6}{1cm}{Error\\Gains}} &       $X_2X$ &    \tabincell{c}{6102.5\\$\pm$826.0}  &    \tabincell{c}{6646.1\\$\pm$585.4}  &     \tabincell{c}{173.2\\$\pm$25.1}  &    \tabincell{c}{2182.3\\$\pm$265.5}  &    \tabincell{c}{2680.2\\$\pm$410.9}  &     \tabincell{c}{159.6\\$\pm$33.7}  &    \tabincell{c}{1050.0\\$\pm$116.7}  &     \tabincell{c}{135.6\\$\pm$16.4}  &      \tabincell{c}{92.9\\$\pm$10.1}  \\
\multicolumn{1}{c|}{} &       $X_1X$ &     676.2$\pm$57.2  &     549.3$\pm$30.3  &     471.9$\pm$43.5  &     674.9$\pm$84.0  &     502.4$\pm$47.6  &     441.9$\pm$34.9  &     572.0$\pm$36.5  &     465.1$\pm$36.9  &     727.7$\pm$61.4  \\
\multicolumn{ 1}{c|}{} &       $X_0X$ &      58.7$\pm$7.4  &     102.5$\pm$11.7  &       NULL &      63.1$\pm$9.8  &      93.8$\pm$10.7  &       NULL  &     115.9$\pm$9.7  &      NULL  &      NULL  \\
\multicolumn{ 1}{c|}{} &       $Z_2Z$ &    \tabincell{c}{7969.1\\$\pm$890.3}  &   \tabincell{c}{55921.2\\$\pm$5978.7}  &  \tabincell{c}{155437.3\\$\pm$15126.6}  &    \tabincell{c}{2284.4\\$\pm$169.8}  &   \tabincell{c}{15283.9\\$\pm$837.0}  &   \tabincell{c}{44795.7\\$\pm$1056.5}  &    \tabincell{c}{5387.8\\$\pm$518.6}  &   \tabincell{c}{13137.3\\$\pm$693.9}  &    \tabincell{c}{2241.6\\$\pm$132.3}  \\
\multicolumn{ 1}{c|}{} &       $Z_1Z$ &     \tabincell{c}{748.5\\$\pm$44.7}  &    \tabincell{c}{3687.2\\$\pm$315.7}  &    \tabincell{c}{5446.5\\$\pm$374.7}  &     \tabincell{c}{677.6\\$\pm$45.1}  &    \tabincell{c}{2290.0\\$\pm$76.8}  &    \tabincell{c}{2785.2\\$\pm$131.7}  &    \tabincell{c}{1930.9\\$\pm$126.5} &    \tabincell{c}{1843.4\\$\pm$46.0}  &    \tabincell{c}{1588.0\\$\pm$71.8}  \\
\multicolumn{ 1}{c|}{} &       $Z_0Z$ &      59.8$\pm$8.3  &     515.9$\pm$24.8  &       NULL  &      63.6$\pm$8.6  &     294.0$\pm$25.4  &       NULL  &     229.2$\pm$19.1  &       NULL  &       NULL  \\
\hline
\multicolumn{ 2}{c|}{$R$} &   \tabincell{c}{7.36E-06\\$\pm$2.67E-07} &   \tabincell{c}{4.15E-05\\$\pm$2.34E-06} &   \tabincell{c}{1.57E-04\\$\pm$7.26E-06} &   \tabincell{c}{9.53E-07\\$\pm$7.58E-08} &   \tabincell{c}{8.90E-06\\$\pm$5.43E-07} &   \tabincell{c}{3.16E-05\\$\pm$2.42E-06} &   \tabincell{c}{1.35E-06\\$\pm$2.60E-07} &   \tabincell{c}{5.36E-06\\$\pm$7.26E-07} &   \tabincell{c}{2.62E-07\\$\pm$9.03E-08} \\
\hline
\hline
\end{tabular}
\end{sidewaystable}
\end{center}

\clearpage

\setcounter{figure}{0}
\setcounter{table}{0}
\setcounter{equation}{0}

\onecolumngrid

\global\long\def\theequation{S\arabic{equation}}
\global\long\def\thefigure{S\arabic{figure}}
\renewcommand{\thetable}{S\arabic{table}}
\renewcommand{\arraystretch}{0.6}

\normalsize

\vspace{1.0cm}


\end{document}